# AN UNSUPERVISED MACHINE LEARNING APPROACH FOR GROUND-MOTION SPECTRA CLUSTERING AND SELECTION

**R. Bailey Bond**[a], **Pu Ren**[b], **Jerome F. Hajjar**[a,1] **and Hao Sun**[c]

[a] Department of Civil and Environmental Engineering, Northeastern University, Boston, MA 02115, USA

[b] Machine Learning and Analytics Group, Lawrence Berkeley National Lab, Berkeley, CA 94720, USA

[c] Gaoling School of Artificial Intelligence, Renmin University of China, Beijing, 100872, China

Clustering analysis of sequence data continues to address many applications in engineering design, aided with the rapid growth of machine learning in applied science. This paper presents an unsupervised machine learning algorithm to extract defining characteristics of earthquake ground-motion spectra, also called latent features, to aid in ground-motion selection (GMS). In this context, a latent feature is a low-dimensional machine-discovered spectral characteristic learned through nonlinear relationships of a neural network autoencoder. Machine discovered latent features can be combined with traditionally defined intensity measures and clustering can be performed to select a representative subgroup from a large ground-motion suite. The objective of efficient GMS is to choose characteristic records representative of what the structure will probabilistically experience in its lifetime. Three examples are presented to validate this approach, including the use of synthetic and field recorded ground-motion datasets. The presented deep embedding clustering of ground-motion spectra has three main advantages: 1. defining characteristics the represent the sparse spectral content of ground-motions are discovered efficiently through training of the autoencoder, 2. domain knowledge is incorporated into the machine learning framework with conditional variables in the deep embedding scheme, and 3. method exhibits excellent performance when compared to a benchmark seismic hazard analysis.

## INTRODUCTION

Feature engineering is critical in representation learning and is used in many domains with applications like stock price forecasting, neurological diagnosis, and acoustic scene classification [1–3]. Unsupervised machine learning (ML) is a tool in artificial intelligence used to deduce the natural structure within a dataset lacking target labels. Unsupervised feature extraction can enhance the characterization of sequence data with complex frequency content and can be used to select a subgroup from a large dataset. A few existing methods of feature extraction include principal component analysis (PCA), local linear embedding, and autoencoders [5–7]. An autoencoder employs the nonlinear parameter discovery of a neural network (NN) to identify relationships that can outperform the linear dependencies of other existing methods [7]. Furthermore, modifications to the typical autoencoder architecture have been studied to accomplish specific tasks [8].

---

[1] Corresponding Author

The deep embedded clustering scheme is adopted here which is defined by a parameterized nonlinear mapping from the data space to a lower dimensional feature space, where a clustering objective is then optimized [9]. With the low dimensional feature space, different clustering algorithms can be performed to identify groups of similar data. A selection from each cluster can be added to a subgroup, theorized to be representative of the complete original dataset.

This deep embedding selection process can be useful in the context of ground-motion selection (GMS) used for seismic hazard analysis (SHA). The suite of ground-motions selected for loading when conducting nonlinear seismic response analysis for civil structures can incur large computational demands and sizable effects to the predicted structural behavior [10]. Additionally, suites of simulated ground-motions can result in thousands of ground-motion time series [11] that would be impractical to utilize, in its entirety, for multi-objective performance assessments. An optimal selection for structural analysis will include as few ground-motions as possible while still expressing all available and relevant records [12]. Consequently, there are multiple methods of selection used in practice with some divergences to the final structural response analysis [12,13].

Some existing methods of record selection include clustering based on physical attributes such as magnitude, peak ground acceleration, or other intensity measurements (IMs) [14–17]. The generalized conditional intensity measure (GCIM) approach, for example, utilizes random realizations of conditional multivariate distributions of static IMs, conditioned on the occurrence of a single IM, from ground-motions for probabilistic selections and statistical comparisons [17,18]. Although these traditional IMs might possess connections to dynamic properties of earthquakes, this categorical approach might fail to include critical latent attributes resulting in a selection based on static and finite features. Another existing method of GMS is the conditional mean spectra (CMS), where a statistical analysis of the spectrum conditioned on a period of interest reveals a hazard-consistent design spectrum [19,20]. The CMS leads the designer to a more realistic target spectrum compared to the conservative Uniform Hazard Spectrum (UHS) provided by building codes. Spectral matching (shape matching) can then be performed to match the hazard-consistent design spectrum at all spectral periods, specifically the period of interest, by modifying the frequency content of the time-series [21,22]. The CMS method, however, only provides a target spectrum. For example, if a spectrum matches the CMS, then it is viewed as a valid selection. This selection might be ignoring critical spectral characteristics of another spectrum that also matches the CMS but was not chosen. More novel methods of classifying ground-motion features that aim at characterizing the spectral frequency content have been proposed [23]. In this method, $k$-means clustering, and self-organizing map (SOM) network algorithms are used to reveal six descriptive features. However, these features still rely on human engineered parameters that do not incorporate the entire spectral behavior. Nevertheless, the main advantages to these methods are to reduce the number of ground-motion records required for engineering analysis and provide probabilistically relevant spectra to the site of interest.

ML is becoming increasingly common in earthquake engineering and seismological applications [24], specifically as it relates to the application of unsupervised learning and ground-motion clustering [14,15,23,25–28]. These studies emphasize the physical attributes of earthquakes, such as magnitude, source-to-site distance, and other IMs. However, solely relying on static data-based clustering analysis may not provide a comprehensive understanding of the concealed patterns and inherent characteristics of synthetic and measured records. To further address the lack of full representation (e.g., the consideration of the complex spectral representations), ML can encode full spectra into a primary set of characteristics and is used to diversify the selection process. Zhang et al. utilized a modified *k*-means clustering algorithm using ML to classify ground-motions based on their response spectra, accentuating the capability of clustering techniques to uncover hidden patterns in ground-motions [29]. Hu et al. compared many dimensionality reduction ML techniques that result in effective characteristic representations of ground-motions for GMS and efficient seismic fragility analysis [26]. To this end, a unique unsupervised NN autoencoder method to cluster spectra and select ground-motions from a group of available records is presented here. In this process, assumptions of statistical distributions of ground-motion characteristics are left out. The following explains a process of spectral selection by applying a general sequence clustering approach where key features of the spectra will be learned by the deep embedding framework and useful to the selection process.

In this approach, we consider synthetically derived ground-motions as well as recorded ground-motions obtained from the Pacific Earthquake Engineering Research Center (PEER) NGA West 2 Database [30]. A computationally efficient fully connected NN autoencoder is constructed to reveal latent features within the suite of ground-motions, capturing their low-dimensional machine-discovered spectral characteristics. By incorporating these features with known static intensity measures (IMs), we condition the component space and generate more representative selections of ground-motions for hazard analysis. This approach effectively combines domain knowledge of ground-motion intensity with machine-discovered sparse spectral characteristics, thus enhancing the reliability of the method. The presented case studies not only demonstrate correlations between the discovered latent features and traditional ground-motion IMs, providing users with a relative understanding of the interpretability of these features, but also showcase excellent performance in a probabilistic SHA with results that are comparable to conventional GMS methods.

## METHODOLOGY

This section outlines the formulation of the deep embedding unsupervised ML clustering approach for GMS. The autoencoder can be trained on a suite of ground-motion spectra from which a selection is desired, resulting in a user defined number of latent features. Each set of latent features correlates to an input spectrum. A clustering algorithm is used to group the spectra in the discovered latent features and defined IMs space, then a representative selection is made from each cluster.

### Data Selection and Scaling

For NN training, a rich inclusive dataset is desirable. From testing different scaling methods, the authors hypothesize that the autoencoder's ability to learn effective representations correlates to preserving the shape of the spectra; min-max scaling of the natural log of the spectral acceleration was adopted. Because it is desired to select ground-motions from all data available, the spectra were not split into a testing set and training set, rather, all data was used for training. This approach is common for unsupervised ML applications with data lacking target labels, requiring other means of validation. Additionally, spectral series were used over time-series ground accelerations due to their significant influence on the structural response of civil structures and broad use in earthquake engineering analysis and design.

The presented framework can be applied to any group of ground-motions (e.g., a set of physics-based simulated ground-motions, a collection of ground-motions matching statistically significant parameters from a Probabilistic Seismic Hazard Analysis (PSHA), or a group of hazard-consistent spectra, conditioned on a period of interest (e.g., resulting from CMS analysis). To show broad use cases, the manuscript only delineates general examples, which highlight the quality and characteristics of the presented latent space, and the ability to select ground-motions that are sufficiently spread in this space.

**Autoencoder Implementation and Training**

The proposed network architecture is illustrated in Figure 1. Autoencoders were developed to uncover latent features, or lower dimensionality encodings for unlabeled data [7]. Stochastic gradient decent (SGD) via backpropagation on a traditional mean square error (MSE) objective function learns the mapping, which is parameterized with $\theta$ and $\emptyset$ by a deep NN encoder and decoder (e.g., $g_\emptyset(x)$ and $f_\theta(x)$ respectively). The formal objective function for $n$ observations of input sequence $\boldsymbol{x}^{(i)}$ with is shown as follows:

$$\mathcal{L}(\theta, \emptyset) = \frac{1}{n}\sum_{i=1}^{n}\left(\boldsymbol{x}^{(i)} - f_\theta\left(g_\emptyset(\boldsymbol{x}^{(i)})\right)\right)^{\boldsymbol{2}}. \tag{1}$$

where $\emptyset = \{\boldsymbol{W}_\emptyset, \boldsymbol{b}_\emptyset\}$ and $\theta = \{\boldsymbol{W}_\theta, \boldsymbol{b}_\theta\}$ are the model parameters (e.g., weights and biases) for the encoder and decoder, respectively. The autoencoder compresses the input to latent representation using fully connected (FC) nodes and then reconstructs the original signal and minimizes the reconstruction error. For more information pertaining to autoencoders, the authors defer to [8].

To enhance performance of the autoencoder, a residual connection from the first layer in deconstruction to the last layer in reconstruction has been found to guide the latent space to more optimal result, while improving the reconstruction. A tensor concatenation was performed on the copied parameters from the encoder's first hidden layer with the decoder's last hidden layer. This type of construction allows subsequent hidden layers the ability to re-use prior feature representations while providing better gradient propagation that is accomplished with implicit deep supervision [31]. The benefits of the residual connection are two-fold. Firstly, the residual connection leads to a significantly better reconstruction of the spectra as the decoder's formulation of the spectra is an important part of the evaluation of the performance of the autoencoder. Although the reconstructed spectral will not be utilized for clustering and selection, without sufficiently low reconstruction error, the latent space should not be trusted as an accurate representation. Secondly, the latent space possesses more interpretable organization of recognizable features (e.g., static IMs). An ablation study that shows the reconstruction and organization of the latent space without the residual connection is presented for the real-world example to further examine the effects of the residual connection.

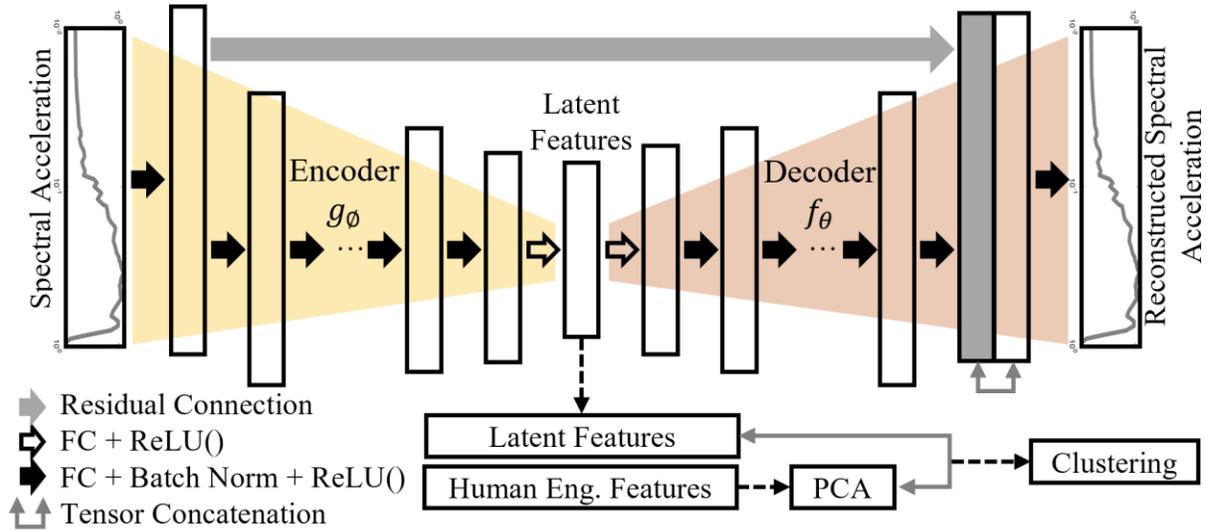

*Figure 1: Network architecture.*

A parametric analysis was conducted to optimize the autoencoder's hyperparameters. The number of fully connected hidden layers and the number of nodes in the hidden layers were generally reduced from the previous dimension by multiples of 2 until reduction approached the user defined number of latent features. The dimension of the spectra is 500 for both the synthetic spectra and field measured spectra and 5 latent features were chosen to sufficiently encode the spectra. Consequently, the hidden layers for the encoder were of the form 500-256-128-64-32-12. The decoder possesses the same structure in reverse. The rectified linear unit (ReLU) activation function was found to outperform other activation functions for the task of reconstruction. Batch normalization (Batch Norm.) layers are used between hidden layers to obtain a steady distribution of activation values during training (e.g., see Figure 1). The minibatch size is set to 50, the weight decay parameter is set to 1e-6, and the starting learning rate is set to 2e-3, with a learning rate decay period of 200 epochs and a multiplicative factor of 0.9. All the above hyperparameters are held constant across all datasets and were found to suitably minimize the reconstruction error in all cases (e.g., see Figure 2). Data specific tuning of these hyperparameters might improve performance on each dataset, but idealistic parameter tuning is impractical for real world implementation.

Other architectures and training methods including additional residual connections, convolutional layers, transfer learning, and a variational latent space will be explored in our future work. Additionally, to incorporate more information in the machine discovered latent features, manifold ground-motion representations (e.g., the spectral displacement, spectral velocity, spectrogram, and ground-motion time-series) can be combined as input/output of the autoencoder. Initial results show including multiple representations of the ground-motion could enhance the latent feature encoding and lead to a more robust GMS method across different datasets.

**Ground-Motion Clustering and Selection**

Once the autoencoder is shown to have sufficient reconstruction ability (e.g., the training loss is sufficiently minimized), one set of latent features is assumed to represent the sparse spectral content of a single ground-motion in the training set. For clustering, a principal component analysis (PCA) is used to condense a set of human engineered features (e.g., various IMs of the ground-motion) into 5 principal components. Together, the set of 5 latent features and 5 human engineered components (termed the *component space*) are used in the clustering algorithms. The component space is equally weighted by traditional IMs, interpretable by the user, and machine discovered latent features resulting from black-box training.

Although the specific clustering method will influence the exact spectra chosen from the component space, if clustering results in spectra that are sufficiently separated in this space, then the exact means of clustering is somewhat trivial. Clustering methods may be chosen by a user based on structures of the component space. Three clustering algorithms are presented to demonstrate the deep embedding and clustering framework (e.g., standard Euclidean $k$-means clustering, hierarchical agglomerative clustering, and a gaussian mixture modeling approach).

For seismic engineering applications, $k$-means clustering, further referred to as KM in this paper, has been a widely used method [15,23,26,27,32–34]. In $k$-means clustering, the principal components of the ground-motions are assembled into $k$ clusters utilizing the vector of components as observations. A set of observations $(\boldsymbol{z}_1, \boldsymbol{z}_2, \dots, \boldsymbol{z}_n)$, where each observation (e.g., a six-dimensional list of components) is partitioned into $k$ clusters $(C_1, C_2, \dots, C_k)$ to minimize the within-cluster sum of squares (WCSS). The formal objective function is shown as follows.

$$\arg \min_C \sum_{i=1}^{k} \sum_{\boldsymbol{z} \in C_i} \|\boldsymbol{z} - \boldsymbol{\mu}_i\|^2 = \arg \min_C \sum_{i=1}^{k} |C_i| \text{Var } C_i \qquad (2)$$

where the $\boldsymbol{\mu}_i$ is the mean of observations in $C_i$. The first step is to define $k$ centroids, choosing them as far away from each other as possible. Next, all remaining observations are associated to the nearest centroid with the smallest Euclidean distance. When all observations are assigned to a group, $k$ new centroids are calculated, and the next iteration is performed. The loop is performed until the centroids do not move to new locations [35]. It is noted that the k-means clustering algorithm possesses some distinct features that might diminish its clustering performance for specific tasks, including a substantial dependence on initial conditions and clusters of varying sizes, densities, and shapes [15,35,36].

The elbow method and silhouette score were both used to define the optimal number of clusters for the $k$-means approach [35]. The prior calculates the within-cluster sum of squares (WCSS) for different values of $k$, choosing the number of clusters where the distortion first starts to converge, visible as an elbow when plotted. A silhouette value ranges from +1 and -1, where a higher value indicates higher cohesion among the clusters. The silhouette value, (Eqn. 3), is found by comparing a measure of similarity of the point $i$ to its own cluster, $\boldsymbol{a}(i)$ (Eqn. 4), and the measure of dissimilarity, $\boldsymbol{b}(i)$ (Eqn. 5), where $N_i$ is the number of observations belonging to cluster $i$ and $d(i, j)$ is the distance between observations $i$ and $j$ in the cluster $C_i$. For most cases, trends display that the fewer clusters chosen leads to higher cohesion between the clusters.

$$s(i) = \frac{\boldsymbol{b}(i) - \boldsymbol{a}(i)}{\max\{\boldsymbol{a}(i), \boldsymbol{b}(i)\}}, if \ C_i > 1 \qquad (3)$$

where:

$$\boldsymbol{a}(i) = \frac{1}{N_i - 1} \sum_{j \in C_i, i \neq j} d(i, j) \qquad (4)$$

and

$$\boldsymbol{b}(i) = \min_{i \neq j} \frac{1}{N_i} \sum_{j \in C_j} d(i, j) \qquad (5)$$

Hierarchical agglomerative clustering (Agg.) has also been used for clustering static earthquake features [28,37]. In hierarchical agglomerative clustering, a "bottom-up" approach, each observation starts as a separate cluster and iteratively merges, using the pairwise distance or dissimilarity between observations and clusters and a given linkage criterion, to the most similar cluster until a single cluster containing all the data is formed [38,39].

A gaussian mixture model (GMM) was also utilized to perform clustering on the component space beginning with random initialization of Gaussian distribution parameters and iteratively updating them using the Expectation-Maximization (EM) algorithm until convergence. This clustering method is a soft clustering algorithm where each observation is assigned a membership probability, indicating the likelihood of it belonging to each cluster verses a definitive clustering classification or hard clustering algorithm [40].

It is noted that for some GMS tasks, the number of clusters to choose is difficult to ascertain. For practical applications, the number of clusters can be selected contingent on the desired number of spectra for structural analysis and influenced by standards or the available computing power. For selection, centroids of the component space clusters were defined and the spectra "closest" to the centroid was selected as the representative ground-motion for that group.

## EXAMPLES

This section illustrates the deep embedding clustering framework for three examples trained on different datasets, including spectra from synthetically derived and field measured ground-motions. The first example shows the effectiveness of the methodology and substantiates the framework. To the author's best knowledge, earthquake spectra have never been reconstructed with an autoencoder and must be tested on a basic level. The second example compares the performance of the presented selection framework to other GMS strategies through a hazard analysis expanding on work by Bradley et al. [41]. The third example shows a real-world application of spectral latent space representation that resembles a wide range of measured spectral characteristics downloaded from the PEER database [30]. The network training loss for each of the following examples are shown in Figure 2. Additionally, the reconstruction difference between the input and output spectra is shown in Figure 3.

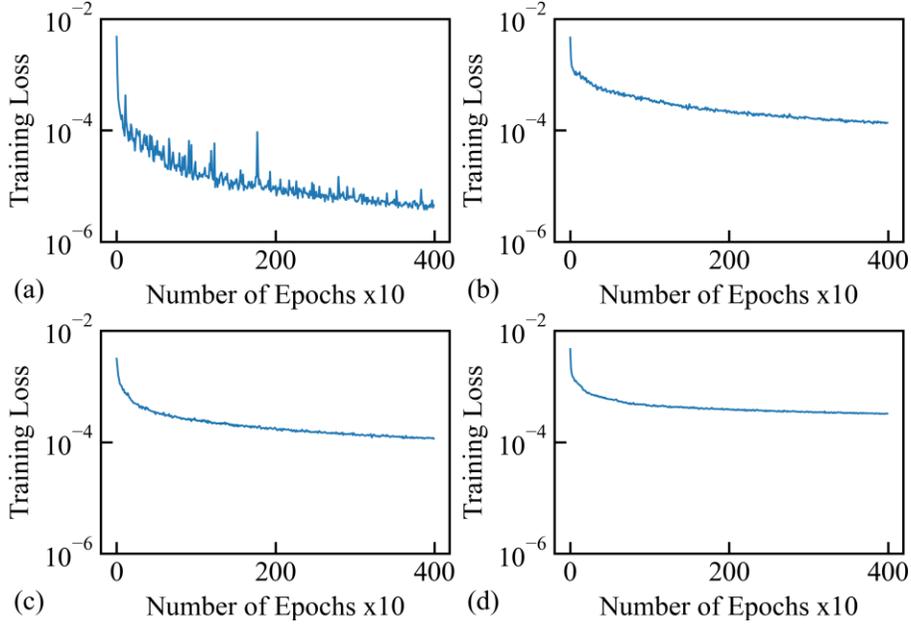

*Figure 2: Network training loss for (a) the synthetic dataset; (b) typical simulation-based SHA example.; (c) the comprehensive PEER dataset. (d) the comprehensive PEER dataset ablation study*

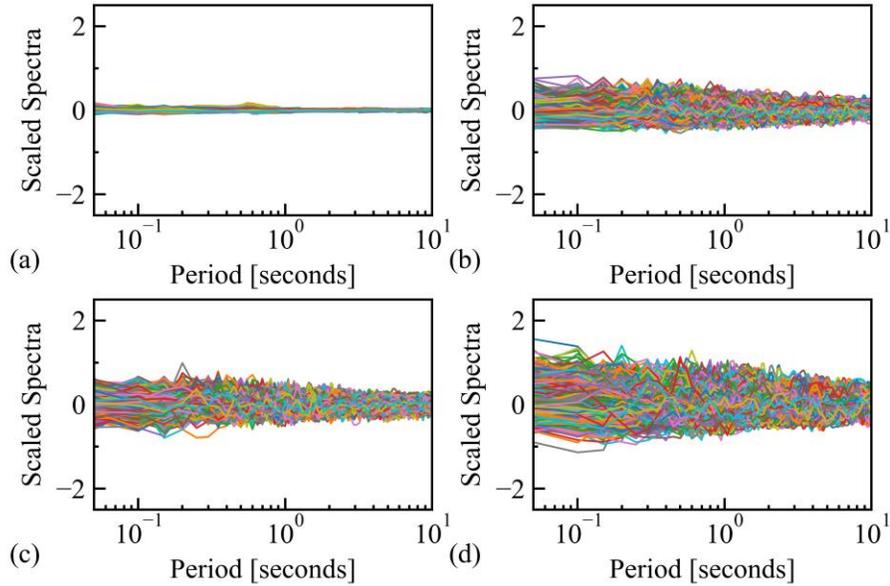

*Figure 3: Network reconstruction difference between input and output spectra for the (a) the synthetic dataset; (b) typical simulation-based SHA example.; (c) the comprehensive PEER dataset; (d) the comprehensive PEER dataset ablation study*

## Clustering Sinusoidal Shape Time-Series

To assess the performance of the autoencoder and clustering framework, 1000 synthetically derived sinusoidal-shape time-series were tested on the framework. This example is a 'toy problem' that supports the application of an autoencoder to embed spectral features to a latent representation. The separation in the latent representation is shown to mimic the separation in the spectral shapes. This example is used to validate the idea that the resulting component space has a correlation to the original spectra.

Five predefined clusters, consisting of 200 sequences each, were generated from sine waves with random phases and magnitudes. Each synthetic cluster possessed additional magnitude oscillations due to added noise and was converted to spectrum with periods ranging from T=0.02 to T=10 with at a 0.02 step size using the OpenSeismo toolbox in MATLAB R2020b [42]. The five time-series clusters and spectral clusters can be clearly identified in Figure 4.

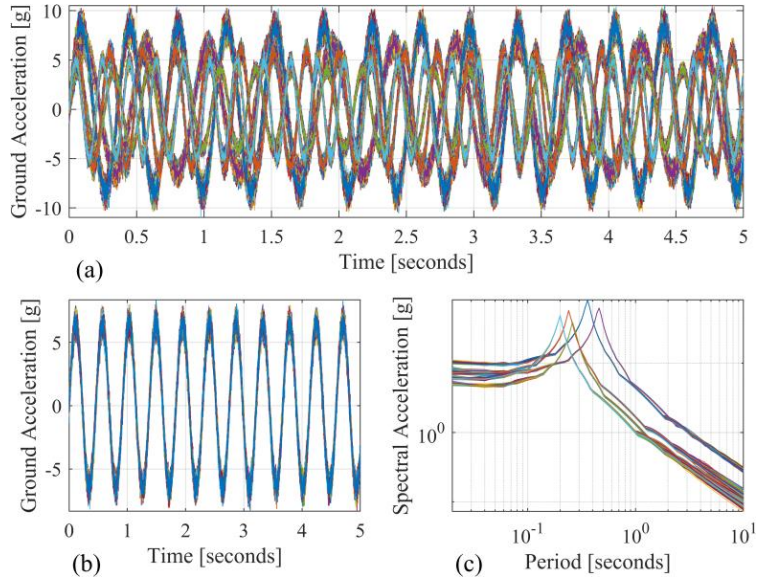

*Figure 4: (a) Five time-series clusters in the synthetic dataset; (b) zoom in view of one cluster of the synthetic dataset; (c) all spectral accelerations of the synthetic dataset*

The proposed framework was used to first discover 5 latent features for each of the 1000 synthetic spectral series. The difference between the reconstruction and original spectral accelerations is shown in Figure 3, verifying the performance of the autoencoder. The human engineered features (or IMs) used in this example are as follows: peak ground acceleration (PGA), peak ground velocity (PGV), peak ground displacement (PGD), Acceleration Spectrum Intensity (ASI), Velocity Spectrum Intensity (VSI), Displacement Spectrum Intensity (DSI), Arias Intensity (AI), 5% - 75% significant duration (Ds575), 5% - 95% significant duration (Ds595), and Cumulative Absolute Velocity (CAV). A PCA is used to linearly combine the calculated human engineered features into 5 defining principal components. Together the 5 IM components and 5 latent features are combined, and clustering is performed on the 10 features directly. To visually investigate a representation of this 10-feature space, an additional PCA is performed to reduce the 10 features into 3 components, termed the component space, and plotted on a 3D graph, shown in Figure 5. Plots of the component space are used for visual purposes only, and clustering is performed on the unreduced set of 10 features (e.g., 5 latent features and 5 IM components).

It was found that the deep embedding framework, with all three clustering algorithms and a predefinition of five clusters (e.g., $k = 5$), can accurately identify ground truth clusters with 100% accuracy for this simple case.

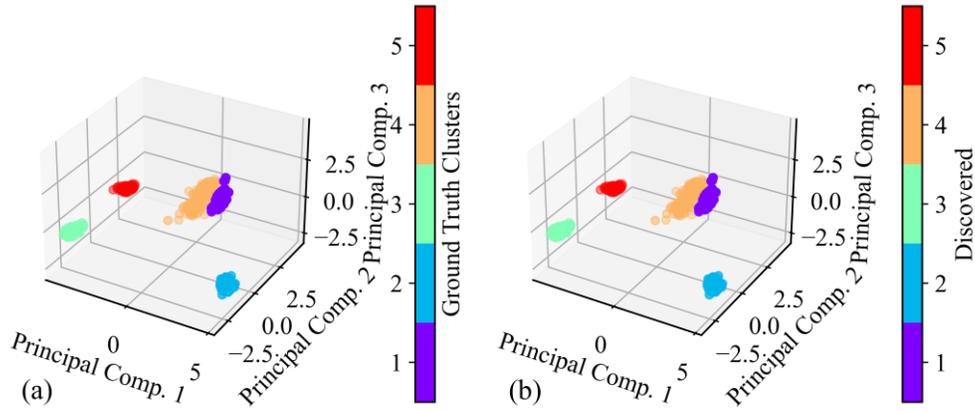

*Figure 5: (a) Ground truth clusters component space depicted with principal components; (b) discovered clusters component space depicted with principal components.*

The silhouette score, a measure of cohesion between members of the same cluster, and the elbow method, a measure of distortion for different defined numbers of clusters, are combined to address the issue of predefining the optimal number of clusters [35]. As shown in Figure 6, the optimal number of clusters using both methods is $k = 5$, where the modeling converges and adding another cluster does not yield a lower distortion in the elbow method or higher cohesion for the silhouette score.

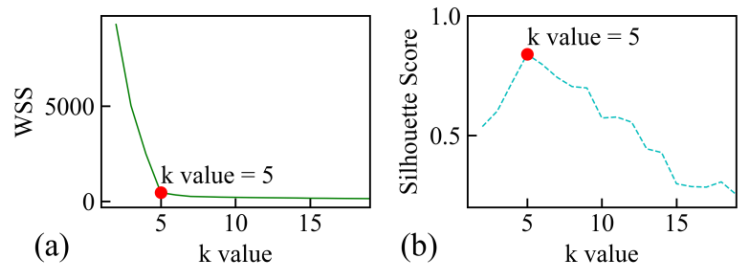

*Figure 6: Identification of optimal number of clusters with the (a) elbow method and (b) the silhouette score.*

### Simulation-Based Ground-Motion Selection for Seismic Hazard Analysis

In this example, a comparison of the deep embedding clustering and selection framework is made to traditional GMS methods expanding on work by Bradley et al. (2015). In Bradley's study, 10 independent sets of 5,000 ground-motions are used in a simulation-based SHA [41]. The case study utilizes the response of a simplified structure at a site in Canterbury, New Zealand with a specific earthquake rupture forecast (ERF) [43] to show how various GMS methods impact SHA results. A 'benchmark' case, termed the direct analysis method or method 1, analyzes the full structural response for all ground-motions in each set, creating a probabilistic seismic hazard curve (SHC). It should be noted that this method is computationally intensive and impractical for most seismic analysis applications. The 'benchmark' case is compared to three traditional GMS methods (e.g., stratified sampling, simple multiple stripes, and GCIM-based selection) which are further explained in Bradley's 2015 paper [41].

Similar to the original paper [41], SHCs are constructed for the presented autoencoder clustering and selection method (termed the AE method) and compared to the 'benchmark' case to analyze the impact of this selection methods on seismic demands. Additionally, comparisons are made to the traditional GMS methodologies in Bradley's work.

In the 'direct analysis' method, all simulated ground-motions that are used to obtain the SHC are used in the seismic response analysis. In the other GMS methods, only selected ground-motions are used in seismic response analysis and weighting parameters are set to adjust exceedance rates of the hazard curves accordingly. For example, the annual rate of exceeding an engineering demand parameter (EDP) (e.g., peak displacement) for the 'direct analysis' method is shown in (Eqn. 6). Figure 7 (a) shows 10 SHCs, each representing one set of 5,000 ground-motions of the 10 independent trails and Figure 7 (b) shows the mean and 80% confidence interval of all trials. In method 2 (e.g., stratified sampling) a discrete set of $n$ mutually exclusive bins is made from a selected IM that quantifies the ground-motion severity. For comparison, the IM chosen to condition method 2 was the spectral acceleration at T=1.32 that showed the highest efficiency replicating the mean SHC from method 1. It is noted that the conditioning IM used to create the bins has a significant impact on the performance of method 2 [41]. Within each IM bin, $m$ ground-motions are selected at random. Consequently, the seismic responses from $m$ = 10 ground-motions over $n$ = 10 different IM bins would result in a subset selection of 100 ground-motions from the initial group of 5,000 ground-motions. Method 2 constructs the SHC with (Eqn. 7) by reweighting the annual rate of occurrence associated with each of the $m$ ground-motions and $n$ bins according to (Eqn. 8) [41]. The calculations of annual rate of exceedance for methods 3 and 4 can be found in the original paper and are left out here for brevity [41].

$$\lambda_{EDP}(edp) = \sum_{i=1}^{N_{rup}^*} I(EDP > edp \, |rup_i) * \lambda_i \qquad (6)$$

Where $EDP$ a specific level of seismic demand and $\lambda_{EDP}(edp)$ is the annual rate of $edp$ exceeding $EDP$, $N_{rup}^*$ is the total number of these alternatively-defined ruptures, $I(EDP > edp \, |rup_i)$ is the indicator function equal to one if $EDP > edp$ for the $i^{th}$ rupture and zero otherwise, and $\lambda_i$ rupture occurrence rate [41].

$$\lambda_{EDP}(edp) = \sum_{i=1}^{N_{rup}^*} I(EDP > edp \, |rup_i) * \lambda_{i,SS} \qquad (7)$$

where
$$\lambda_{i,SS} = \lambda_i * \frac{\sum all \; \lambda_i \; in \; bin \; n}{\sum all \; selected \; \lambda_i \; in \; bin \; n} \qquad (8)$$

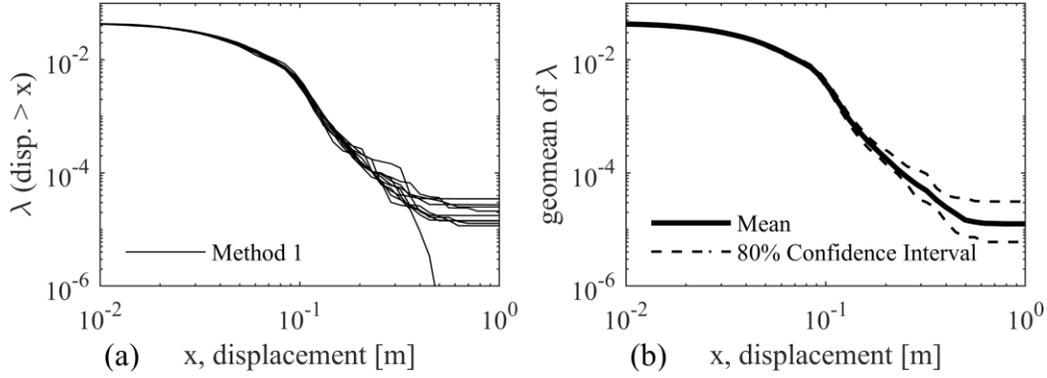

*Figure 7: (a) ten independent sets of SHCs for method 1 [41], (b) the mean and 80% confidence interval of ten independent sets of SHCs for method 1[41]*

The annual rate of exceedance for the AE method presented in this paper can be calculated in a similar fashion to method 2. Instead of bins of ground-motions pertaining to a specific IM, the clustering method devises the component space into clusters and a selection is made from each. In our case, we choose to set the clustering algorithm such that the number of clusters matches the number of desired ground-motions. To match Bradley's paper, 1 selection is made from 100 clusters resulting in a subgroup of 100 ground-motions. The formulation of the annual rate of exceedance for the AE method (e.g., Eqn. (9)) utilizes a weighted variable to adjust the $\lambda_i$ (e.g., Eqn. (10)) for the number of samples associated with each cluster and the corresponding selection.

$$\lambda_{EDP}(edp) = \sum_{i=1}^{N^*_{rup}} I(EDP > edp \,|\, rup_i) * \lambda_{i,AE} \tag{9}$$

where

$$\lambda_{i,AE} = \lambda_i * \frac{\sum all \; \lambda_i \; in \; cluster \; n}{\sum all \; selected \; \lambda_i \; in \; cluster \; n} \tag{10}$$

Comparisons for GMS methods 2, 3, and 4, along with three different clustering algorithms used in the deep embedding autoencoder approach are made with the 'benchmark' results from method 1 (e.g., Figure 8). In all cases, the 80% confidence interval of the GMS method encapsulates the mean exceedance curve of method 1. It is shown that the presented AE approach better replicates the 'benchmark' case, regardless of clustering algorithm, compared to the traditional methods. The $k$-means clustering algorithm matches the mean and confidence interval of method 1 reasonably well and is therefore singled out for further evaluation. Figure 9 compares various metrics of the AE method with the other GMS methods. The component space sufficiently distributes ground-motion records resulting in representative selections when made from this space. To further demonstrate the import of the component space, a 3D example of this space from the first set of 1,000 ground-motions is represented in Figure 10, showcasing an organization of the space from various intensity measures.

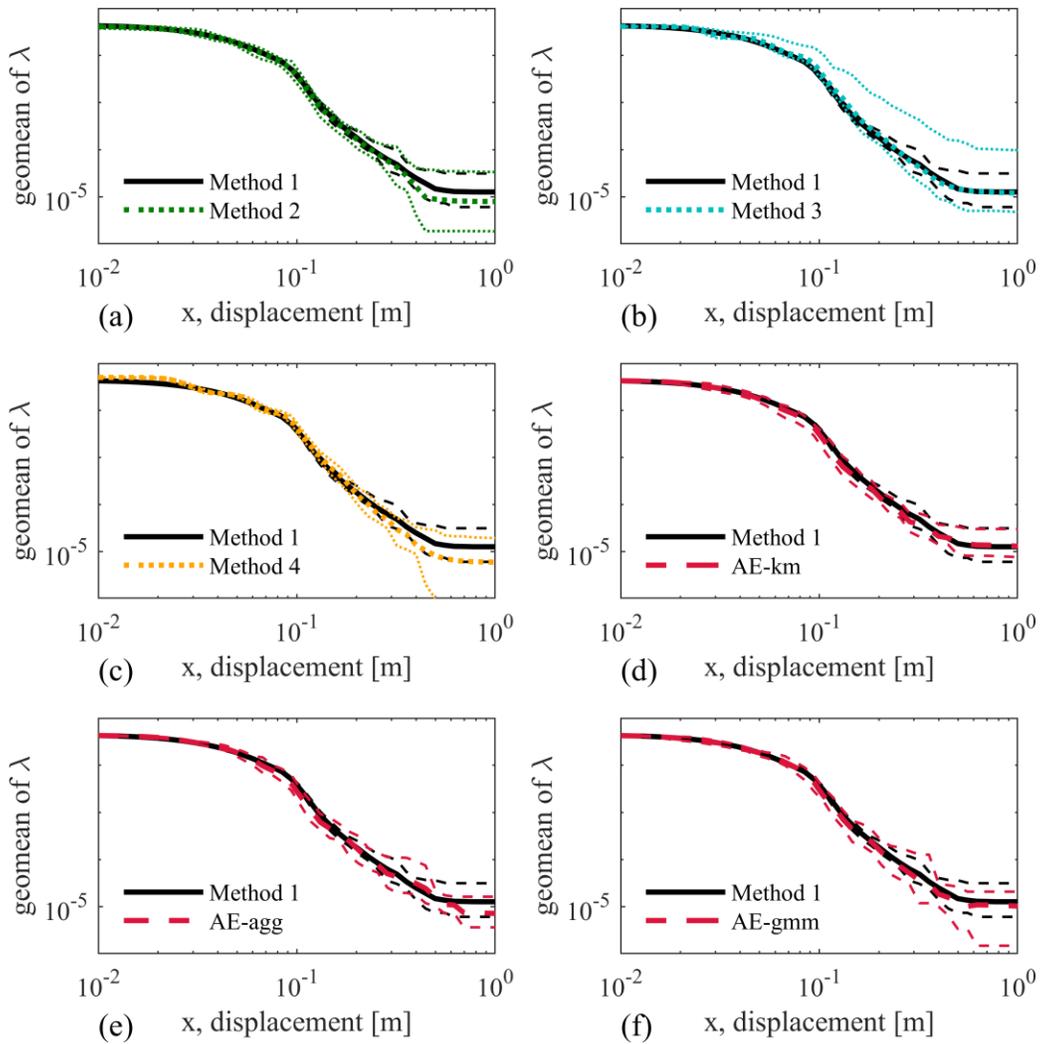

*Figure 8: the mean and 80% confidence interval of ten independent sets of SHCs comparing method 1 to (a) method 2; (b) method 3; (c) method 4; (d) AE method with k-means clustering; (e) AE method with Agg. Clustering; and (f) AE method with GMM clustering*

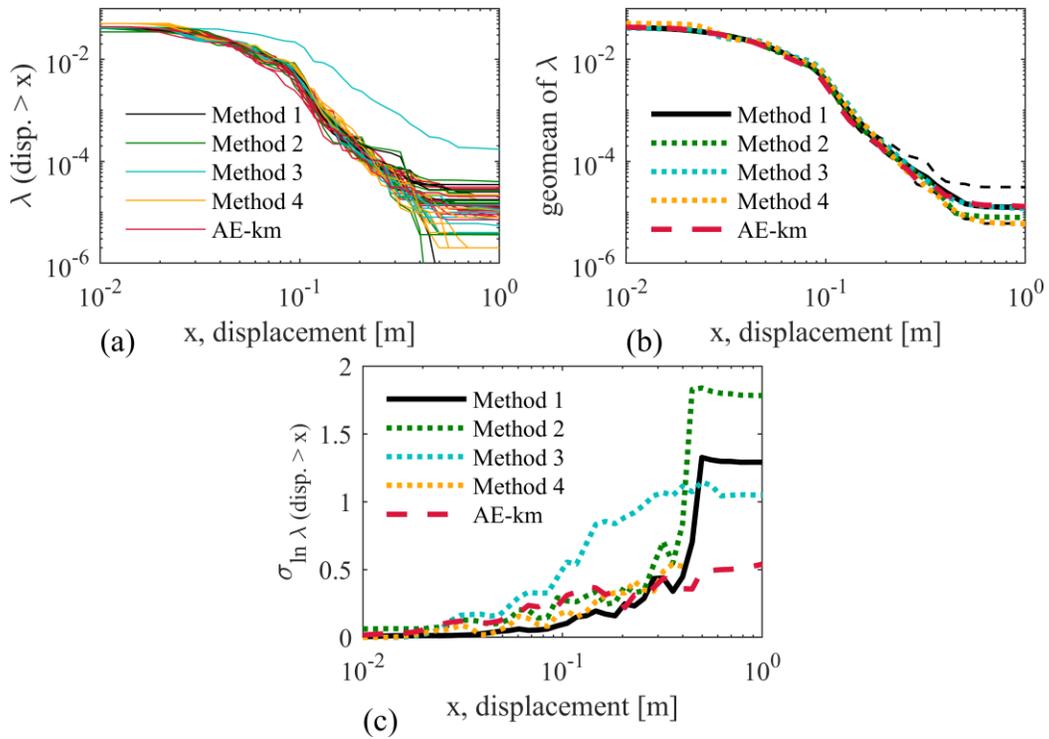

*Figure 9: Comparing methods 1-4 with the AE method with k-means clustering with (a) all 10 independent SHCs for each method;(b) means of SHCs for each method; (c) standard deviation of SHCs for each method.*

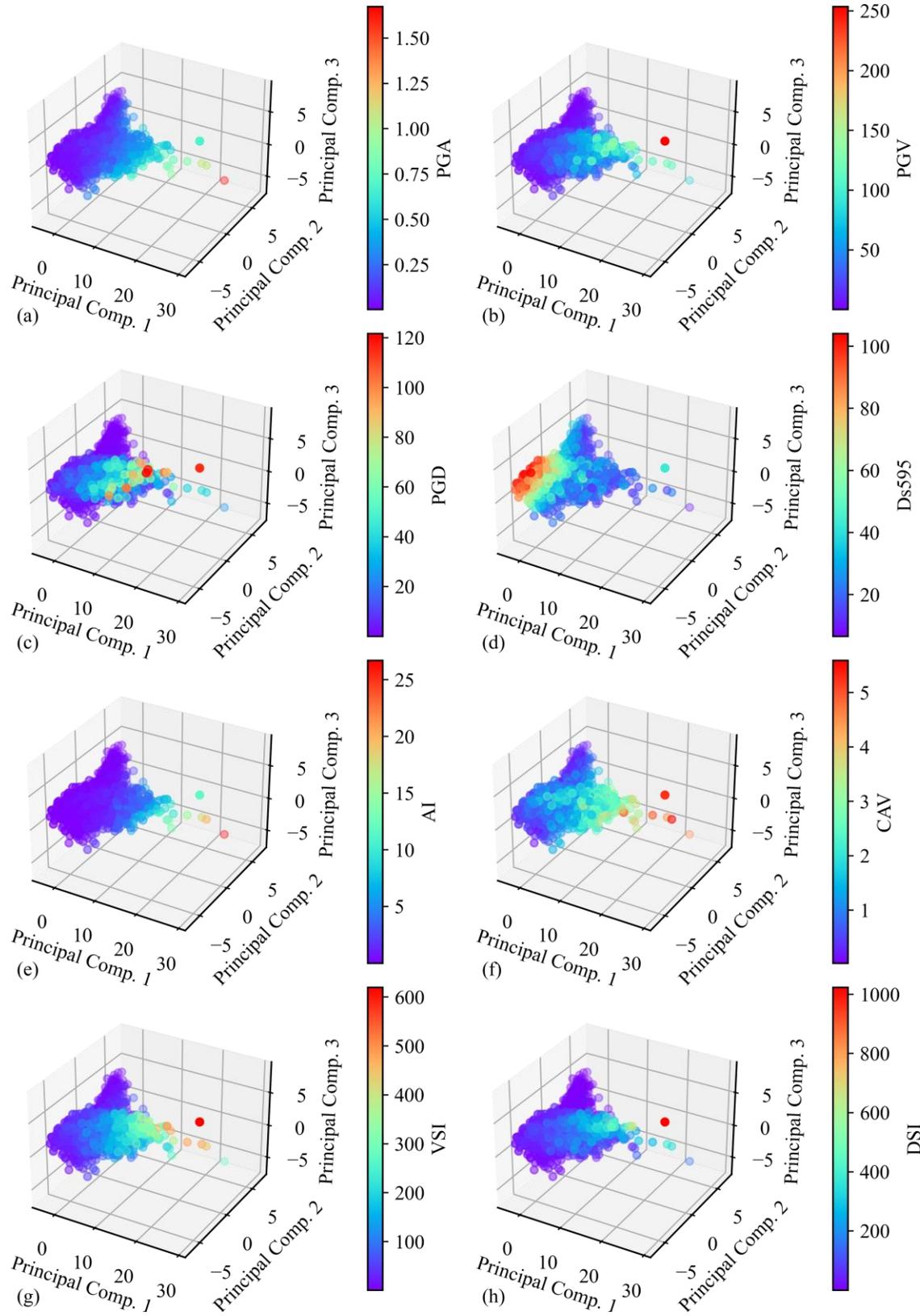

*Figure 10: Component space for Simulation-Based GMS for SHA colored by (a) the PGA; (b) the PGV; (c) the PGD; (d) the 5%-95% significant duration (Ds595); (e) the Arias Intensity (AI); (f) the cumulative absolute velocity (CAV); (g) the velocity spectrum intensity (VSI);, and the displacement spectrum intensity (DSI).*

Three metrics are utilized to display the difference between the comparative GMS methods with the 'benchmark' case and shown in Figure 11. The trapezoidal area between the mean of all trial exceedance curves provides a measure of the energy of the hazard level and quick way to visually compare the two curves but does not consider specific design parameters or probability of exceedance at each value. The root mean squared error (RMSE) (Eqn. (11)) provides a quick and interpretable error metric for the difference of the 'benchmark', $\lambda^1$, and subsampled SHCs, $\hat{\lambda}^\alpha$ where $\alpha = \{method\ 2, method\ 3, method\ 4, agg, km, gmm\}$. The weighted mean square error (WMSE) (Eqn. (12)) is used to give more weight to higher engineering design parameters, which are more likely to cause structural damage and may be more important for assessing the safety of critical structures. The squared error in the WMSE penalizes larger differences more heavily, which may be desirable if larger errors are considered more significant.

$$RMSE = \frac{1}{n}\sum_{i=1}^{n}\left(\sqrt{\left(\lambda_i^1 - \hat{\lambda}_i^\alpha\right)^2}\right) \tag{11}$$

$$WMSE = \frac{1}{n}\sum_{i=1}^{n}\left(\frac{x_i}{\sum x}\left(\lambda_i^1 - \hat{\lambda}_i^\alpha\right)^2\right) \tag{12}$$

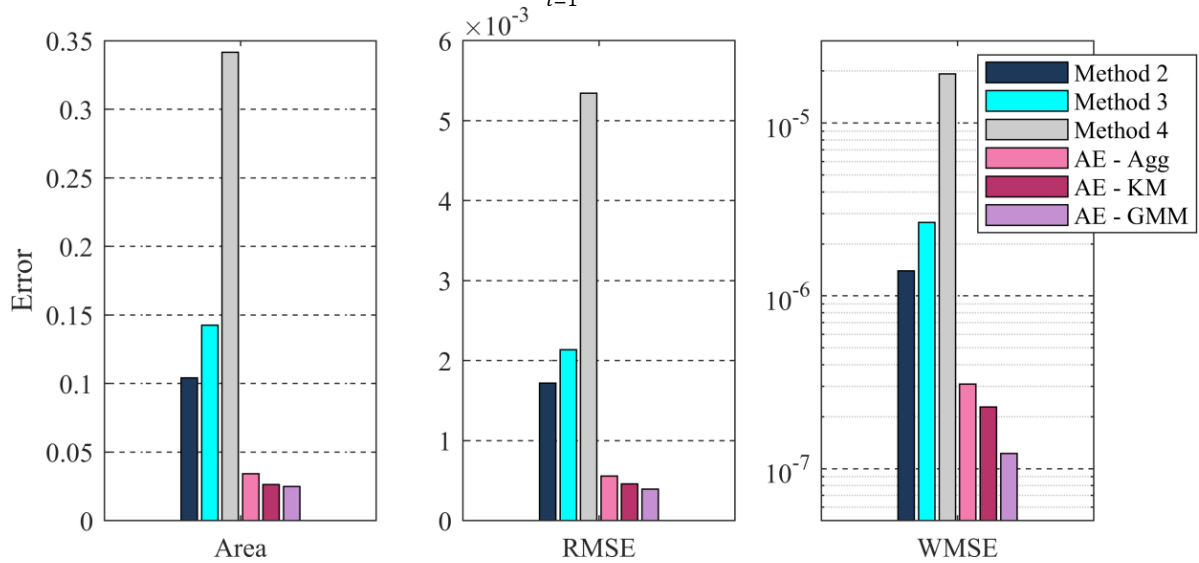

*Figure 11: Error metrics for simulation-based GMS methods for SHA*

**Clustering Earthquake Spectra for Ground-Motion Selection**

The third dataset includes spectral records downloaded from the flat-files from PEER NGA-West 2 Database [30]. Training the network on a comprehensive suite of spectra shows the ability of the latent space to separate features from a complete range of magnitudes and energy contents, practical for real-world applications. In this example, there are no prior assumptions of a site, performance objectives, or target IMs. However, design goals can be implemented by focusing the original dataset within the bounds of a user defined SHA and input to the presented framework.

To match the increment of the prior examples, spectra in this example were interpolated to obtain an even step size of 0.02 with a range of T = 0.02s to T = 10s from the original PEER data with a spline-based interpolation method. This resulted in a final sequence length of 500 for each spectrum. Filters were implemented to only include records with spectral data with a maximum acceleration above 0.1g. This set included 6,600 spectra.

Matching the previous example, each record was encoded to 5 latent features. The static IMs for this example comprised the highest reporting rate from the flat-file can be enumerated as follows: Peak Ground Acceleration (PGA), Peak Ground Velocity (PGV), Peak Ground Displacement (PGD), Magnitude (Mag.), Joner-Boore Distance (Rjb), Shear wave velocity (VS30), Fault Rupture Area (FRA). A PCA was conducted to linearly decompose the 8 static IMs into the 5 principal components that were then combined with the 5 latent features and clustering was performed on the 10-component space. To visualize this space, Figure 12 shows the component space representations of a PCA of the 5 principal components and 5 latent features. The component space is colored by each of the IMs to show the correlation with understandable intensity characteristics.

In practical applications, the desired number of clusters is not always explicitly identified with standard methods (e.g., the elbow method or silhouette method). One method of choosing the number of clusters could be to set the number of clusters to the number of desired spectra, limited by the availability of computational capacity for structural analysis or demands by engineering codes and standards. Multiple spectra could also be chosen from each cluster. For example, one selection could be the spectra closest to the centroidal spectra, and another selection could be the spectra furthest from the centroidal spectra. For this case, the component space was separated into 5 clusters through $k$-means clustering in Figure 12 for a selection of 5 ground-motions to align with Chapter 21 of ASCE 7 (ASCE 7-16: 2000). To further illustrate the clustering result, the scaled spectra is grouped by the assigned clusters from the component space grouping presented in Figure 13 and Figure 14. These images reveal a clear pattern between clustered spectra (e.g., spectral shape). Overall, GMS is accomplished with this framework through interpretable latent features and the flexibility to find clear cluster partitions in the source data.

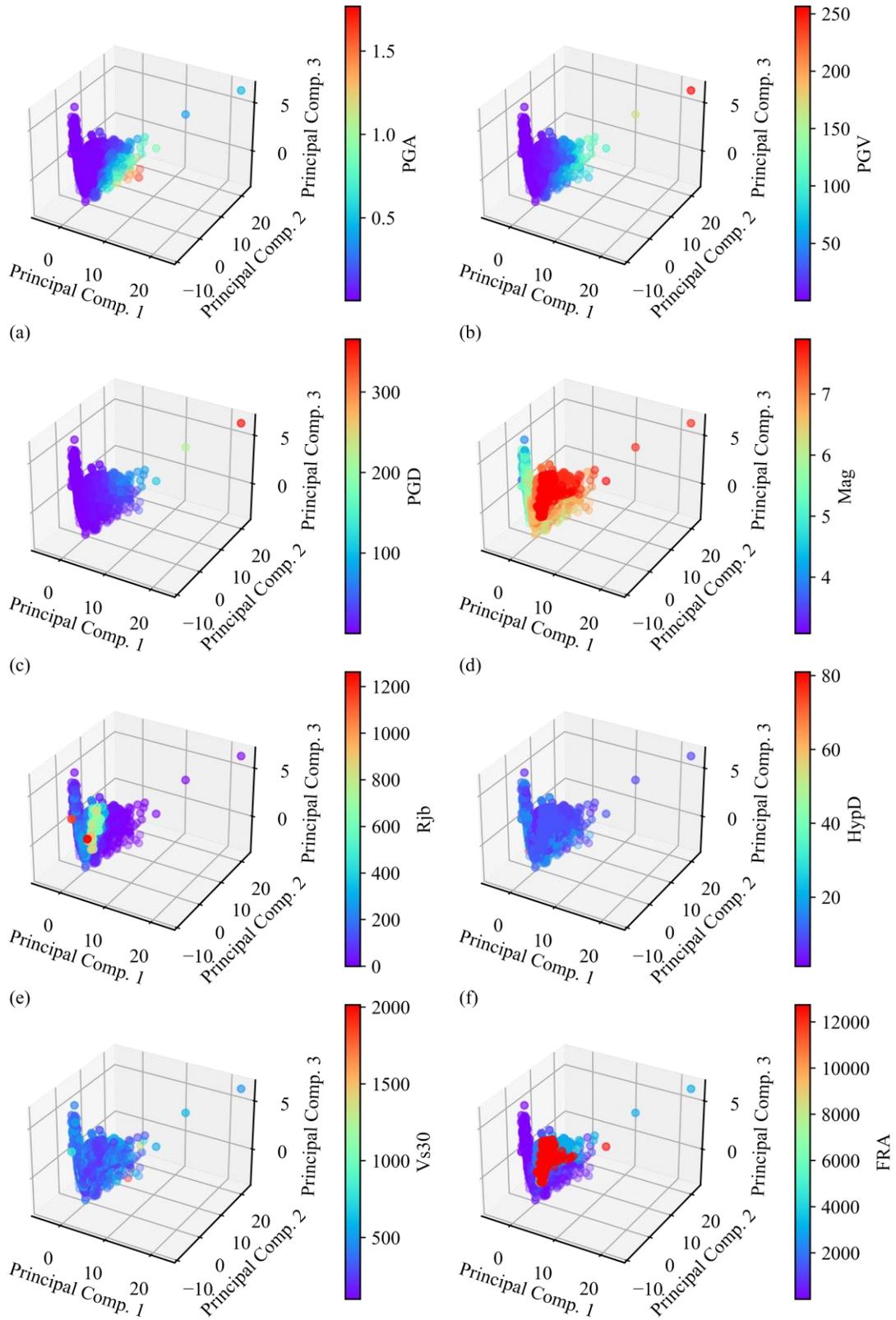

*Figure 12: Component space for comprehensive suite of ground-motions colored by (a) PGA; (b) PGV; (c) PGD; (d) magnitude (Mag); (e) Joyner-Boore source-to-site distance (Rjb); (f) the hypo center depth (HypD); (g) shear wave velocity (Vs30); (h) the fault rupture area (FRA).*

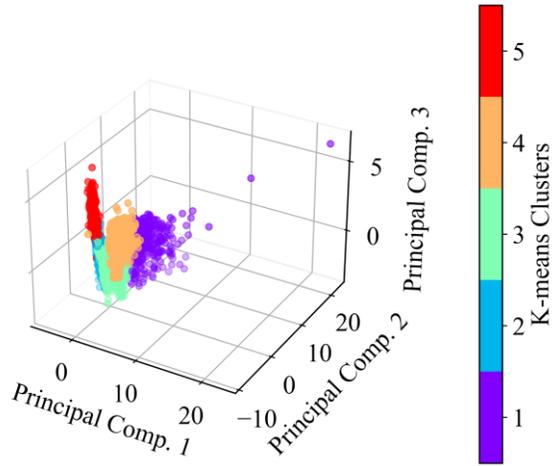

*Figure 13: Component space for comprehensive suite of ground-motions colored by colored by the discovered clusters using k-means*

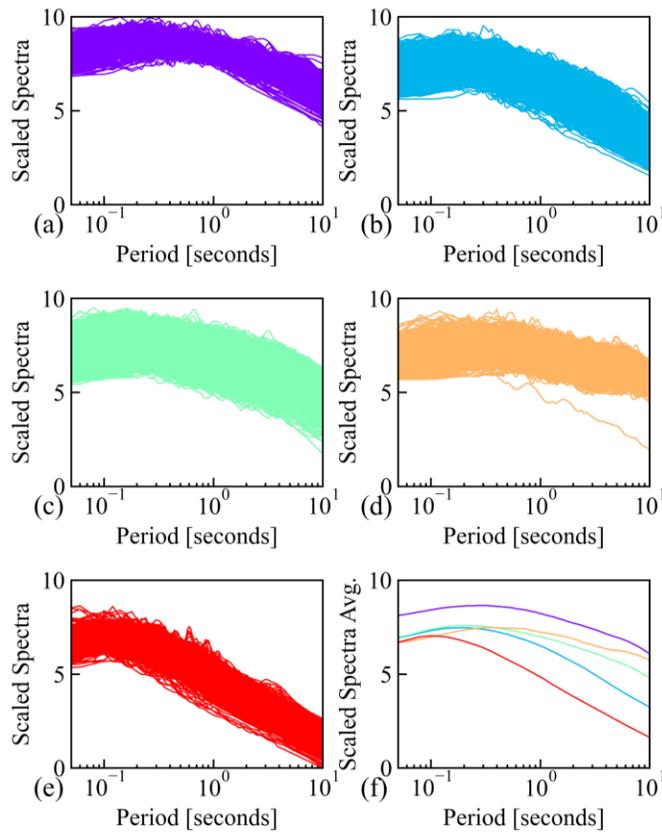

*Figure 14: Clustering of scaled spectra through the proposed framework for (a) cluster 1; (b) cluster 2; (c) cluster 3; (d) cluster 4; (e) cluster 5; (f) average of each cluster*

To deepen the understanding of the presented framework, an ablation study is presented showing the results of the deep embedding without the residual connection. The ablation study network architecture would remove the residual connection from Figure 1. Figure 15 shows the component space of the encoding without the residual connection. Although this space is still organized by the record magnitude, it does not have as clear levels of differentiation compared to the network with the residual connection (e.g., Figure 12 (d)). Figure 16 (a) and (b) show three individual spectral reconstructions for the full network and ablation study, respectively. The inclusion of the residual connection drastically improves the reconstruction performance and organizes the latent feature space into a more interpretable structure.

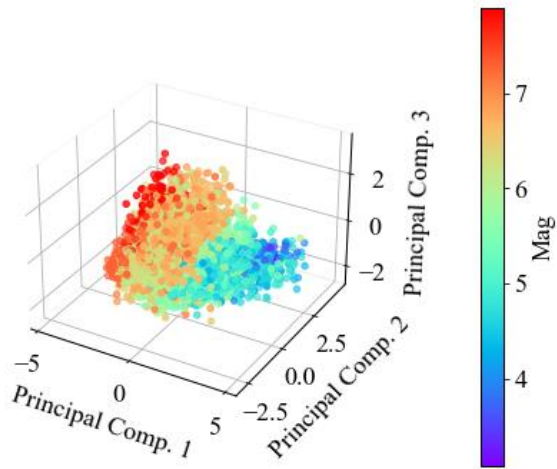

*Figure 15: The component space of the ablation study*

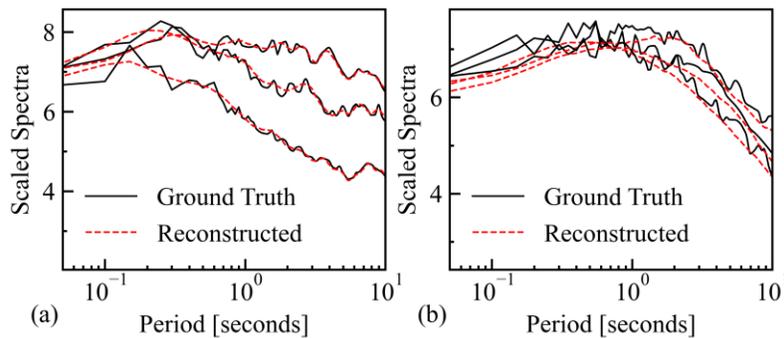

*Figure 16: (a) typical reconstruction for the full network architecture; (b) typical reconstruction for the ablation study.*

## CONCLUSION

This paper presents a deep embedding unsupervised ML approach for GMS. The selection of ground-motion records has a significant role in SHA and surrogate metamodeling of structural systems under earthquake hazards. The proposed framework can uncover latent characteristics of earthquake spectra to expand the overall representation during GMS for nonlinear structural time-history analysis. This work utilized synthetic data and field ground-motion measurements to illustrate the effectiveness of the proposed framework.

The methodology particularly showed excellent clustering ability in the first example, correctly clustering all 1000 synthetic ground-motion spectra by the original groups. In a probabilistic SHA, the AE method produced ground-motions subgroups that exhibit excellent alignment to a 'benchmark' case for multiple error metrics. Additionally, the component space is shown to have high correlation to known physical intensity measures on a suite of measured ground-motions. In general, the framework can uncover latent attributes that maintain interpretable characteristics and are practical for clustering. Furthermore, the proposed approach is fundamental in nature and can be applied to cluster any type of sequential data in the time, frequency, and spectral domains.

## ACKNOWLEDGMENTS

The material presented in this paper is based upon work supported by the National Science Foundation under Grant No. CMMI-2013067 and Northeastern University. Any opinions, findings, and conclusions or recommendations expressed in this material are those of the authors and do not necessarily reflect the views of the National Science Foundation or other sponsors.

## DATA AVAILABILITY STATEMENT

All the source codes to reproduce the results in this study are available on GitHub at https://github.com (detailed URL will be provided after official publication of this paper).